\documentclass{article}
\usepackage{graphicx} 
\usepackage[utf8]{inputenc}
\usepackage[english]{babel}
\usepackage[letterpaper,top=2cm,bottom=2cm,left=3cm,right=3cm,marginparwidth=1.75cm]{geometry}

\usepackage{romannum}
\usepackage{amsmath}
\usepackage{xcolor}
\usepackage{graphicx}
\usepackage[colorlinks=true, allcolors=blue]{hyperref}
\usepackage{amsmath,amssymb}
\usepackage{epsfig}
\usepackage{caption}
\usepackage{epstopdf}
\usepackage{lipsum}
\usepackage{wrapfig}
\usepackage{lmodern}
\usepackage{cite} 
\usepackage{cuted}
\usepackage[colorlinks=true, allcolors=blue]{hyperref}
\usepackage{textgreek}
\usepackage{authblk}
\usepackage{float}
\usepackage[justification=raggedright, singlelinecheck=false]{caption}
\date{}
\title{Active control of phase matching in nonlinear metasurfaces using Pancharatnam--Berry phase}
\author[1]{Madona Mekhael}
\author[2]{Roman Calpe}
\author[2]{Tommi K. Hakala}
\author[1]{Robert Fickler}
\author[1]{Mikko J. Huttunen}

\affil[1]{Photonics Laboratory, Physics Unit, Tampere University, Tampere, Finland}
\affil[2]{Center for Photonics Sciences, University of Eastern Finland, Joensuu, Finland}
\begin{document}

\maketitle
\renewcommand{\abstractname}{}  
\begin{abstract}
Reconfiguring the spectral output of nonlinear metasurfaces after fabrication remains challenging. We address this by exploiting the nonlinear Pancharatnam--Berry phase of $C_{3v}$-symmetric plasmonic metasurfaces. By integrating two metasurfaces inside a multipass cell, we experimentally demonstrate continuous spectral tuning of second-harmonic generation (SHG) phase-matching peaks across a 900--970~nm pump range by rotating one metasurface relative to the other. The extracted geometric phase follows the $3\sigma\theta$ dependence, and a full $2\pi$ tuning cycle is completed with $120^{\circ}$ of physical rotation. This establishes geometric-phase metasurfaces as a reconfigurable nonlinear platform, where mechanical rotation enables post-fabrication and broadband tuning of nonlinear optical responses.
\end{abstract}
\section{Introduction}

The ability to tailor nonlinear optical responses at the sub-wavelength scale has emerged as a cornerstone of modern nanophotonics, offering a flat alternative to traditional bulk nonlinear crystals. Unlike bulk media, which require long interaction lengths for efficient nonlinear processes~\cite{Giordmaine1962PRL, Maker1962PRL}, optical metasurfaces enable nonlinear phenomena within a sub-wavelength thickness by enhancing light--matter interactions through engineered resonances, resulting in relaxed phase-matching requirements~\cite{Kauranen2012NatPhoton, Butet2015ACSNano, Lee2014Nature, Almeida2016NatComm}. 
A defining advantage of metasurfaces is their ability to exert precise control over the local phase of an optical signal, enabling transformative applications in the linear regime such as flat lensing~\cite{khorasaninejad2016metalenses,Chen2012NatCommun,Khorasaninejad2017Science}, holography~\cite{Huang2013NatCommun,Wan2016ACSNano,Ni2013NatComm,Zheng2015NatNanotechnol}, and arbitrary beam shaping~\cite{Kang2012OpticsExpress,Yu2014NatMater,Genevet2017Optica}. Nonlinear metasurfaces extend these capabilities into the frequency-conversion regime, enabling sophisticated wavefront engineering of the generated signal~\cite{Ye2016NatCommun,Gao2018NanoLett,Hahnel2023ACSPhotonics,Karepov2025NanoLett}. Among the various mechanisms for phase control, the Pancharatnam-Berry (PB) phase---also known as the geometric phase---has proven exceptionally robust. Because the PB phase is determined entirely by the spatial orientation of the meta-atoms, it provides a deterministic and broadband method for phase manipulation~\cite{Berry1984ProcRSocA,Bomzon2002OL,Huang2012NanoLett}.   
 
Despite these advances in wavefront control, the vast majority of nonlinear metasurfaces are static: once fabricated, their phase profiles and resulting emission characteristics are fixed. Achieving dynamic tunability---the ability to reconfigure the nonlinear response in real time---remains a significant challenge in the field. Existing approaches rely on external stimuli such as thermal heating~\cite{Rocco2021OpticsExpress} or electrical gating~\cite{Feinstein2025npjNanophotonics,He2024NanoLett,Yu2022NatPhotonics}, which can be slow, power-consuming, or complex to integrate with compact photonic platforms. An alternative approach that has remained unexplored is to exploit the inherent rotational sensitivity of the PB phase: mechanically adjusting the relative orientation between nonlinear elements enables continuous phase modulation without altering the physical properties of the nanostructures.

\begin{figure}
    \centering
    \includegraphics{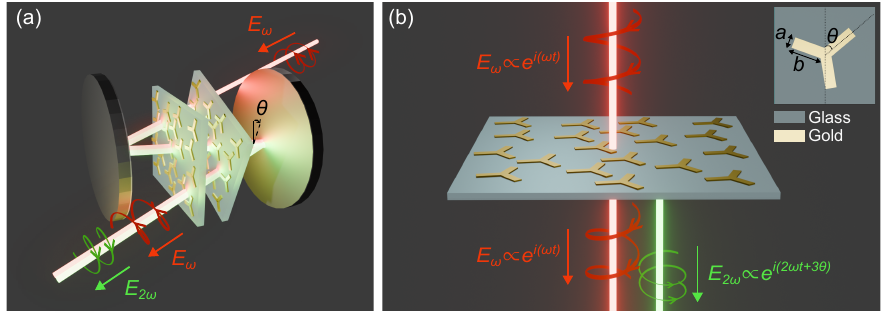}
    \caption{\textbf{Geometric phase control in a multipass cell.} (a) Schematic of the dual-metasurface multipass cell, consisting of two high-reflectivity mirrors enclosing two metasurfaces, one of which is rotated by an angle $\theta$ relative to the other. (b) Schematic of the metasurface illuminated by a right-circularly polarized (RCP) fundamental pump at frequency $E_\omega$. While the transmitted pump remains unmodulated, the generated second harmonic $E_{2\omega}$ is emitted in the left-circular polarization (LCP) state and acquires a nonlinear PB phase determined by the in-plane orientation angle $\theta$ of the meta-atoms. The inset shows a single three-bladed gold nanostructure with arm width $a = 50$~nm and arm length $b = 130$~nm.}
    \label{fig:schematic}
\end{figure}

In this work, we demonstrate a dynamic approach for nonlinear wavefront control by integrating a dual-metasurface configuration within a multipass cell (MPC), schematically presented in Fig.~\ref{fig:schematic}a. One metasurface serves as a stationary reference while a second, mounted on a rotation stage, acts as a tunable phase actuator. Cascaded interactions between the pump beam and both metasurfaces generate phase-matching peaks in the SHG spectra, whose spectral positions are governed by the metasurface geometry. By rotating one metasurface with respect to the other, we exploit the nonlinear geometric phase to introduce a continuously controllable phase delay of up to $2\pi$ into the conversion process, enabling deterministic spectral translation of the SHG phase-matching peaks. This platform offers a versatile and compact route toward reconfigurable nonlinear flat-optics, with potential applications in tunable frequency conversion, nonlinear holography, and on-chip optical signal processing.

\section{Nonlinear Phase-Matching via the Geometric Phase}
In the MPC (Fig.~\ref{fig:schematic}a), the pump interacts repeatedly with the metasurfaces, making the SHG analogous to that of stacked metasurfaces~\cite{Stolt2021PRL}. On each pass, the pump generates a SHG signal that interferes constructively or destructively with the SHG from previous passes, depending on the accumulated phase mismatch between the fundamental and second-harmonic fields. The total SHG intensity is given by:
\begin{equation}
\text{SHG} \propto \left| \sum_{j=1}^{N} T(\omega)^j T(2\omega)^{j/2} e^{\mathrm{i}j\Delta k} \chi_\mathrm{ms}^{(2)} E(\omega)^2 \right|^2,
\label{eq:SHG}
\end{equation}
where $\chi_\mathrm{ms}^{(2)}$ is the relevant component of the nonlinear susceptibility tensor of the metasurface, $N$ is the number of passes, $T(\omega)$ and $T(2\omega)$ are the transmittances at the fundamental and second-harmonic frequencies, and $\Delta k$ is the phase mismatch. When the pump wavelength is scanned, the interference of the SHG signals gives rise to discrete peaks in the SHG spectrum at the phase-matched wavelengths, whose spectral positions are determined by the metasurface design. In the static case, $\Delta k$ depends solely on the propagation phase and the metasurface interaction phase at both the pump and SHG  frequencies~\cite{Mekhael2024ACSPhotonics}.

In the present work, we extend earlier framework by introducing the nonlinear PB phase as an additional tunable control parameter. The PB phase arises when a meta-atom is rotated by an angle $\theta$: for a meta-atom generating the $n$th harmonic under circularly polarized illumination, the emitted signal acquires a spin-dependent geometric phase:
\begin{equation}
\Phi_\mathrm{PB} = (n \pm 1)\sigma\theta,
\label{eq:PB_general}
\end{equation}
where $n$ is the harmonic order, $\sigma = \pm 1$ is the helicity of the incident circular polarization (with $\sigma = +1$ for RCP and $\sigma = -1$ for LCP), and the $+$ and $-$ signs correspond to output light in the opposite ($+$, cross-polarized) and same ($-$, co-polarized) circular polarization state as the input, respectively~\cite{Li2015NatMater}. 
The allowed harmonic orders are governed by the rotational symmetry of the meta-atom~\cite{Chen2014PRL,Li2015NatureMaterials}. A meta-atom with $m$-fold rotational symmetry can only generate harmonics of order:
\begin{equation}
n = jm \pm 1,
\label{eq:selection}
\end{equation}
where $j$ is a positive integer, the sign is $+$ for the co-polarization and $-$ for cross-polarization . 
Adding this contribution, the full phase-matching condition $\Delta k + \Phi_\mathrm{PB} = 2\pi l$ becomes:
\begin{equation}
2(\varphi_\omega + \delta_\omega) - \varphi_{2\omega} - \delta_{2\omega} + \Phi_\mathrm{PB} = 2\pi l,
\label{eq:PM}
\end{equation}
where $l$ is an integer, $\varphi_\omega = k_\omega h$ and $\varphi_{2\omega} = k_{2\omega} h$ are the phase accumulations over a propagation distance $h$, with $k_\omega = 2\pi/\lambda_\omega$ and $k_{2\omega} = 4\pi/\lambda_\omega$ the wavenumbers in air, $\delta_\omega$ and $\delta_{2\omega}$ are the metasurface interaction phases extracted from transmittance spectra. The specific case of meta-atoms with three-fold rotational symmetry, as used in our experiment, are depicted in Fig.~\ref{fig:schematic}b. Since $\Phi_\mathrm{PB}$ depends solely on $\theta$, it acts as a tunable bias that continuously shifts the SHG phase-matching peaks without modifying the material or geometry of either metasurface.

\section{Results and Discussion}

\subsection{Sample design and working principle}
\begin{figure}
    \centering
    \includegraphics{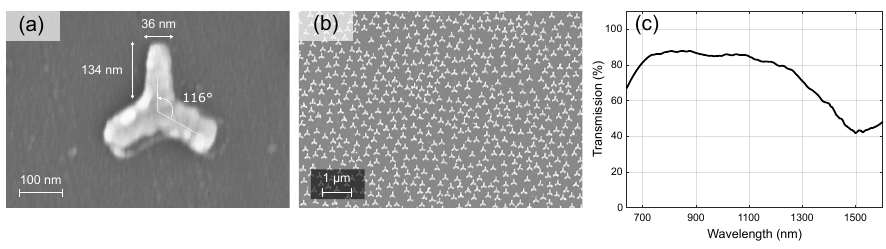}
    \caption{\textbf{Sample characterization.}(a) Scanning electron microscopy (SEM) image of a single 
    fabricated three-bladed gold nanostructure. The measured dimensions are indicated, showing slight deviations from the nominal design values due to fabrication tolerances. (b) SEM image of a larger metasurface area, demonstrating the high fabrication quality and uniformity of the nanostructure array. (c) Measured transmittance spectrum of the metasurface showing plasmon resonance around 1500~nm.}
    \label{fig:fab}
\end{figure}

The metasurface consists of three-bladed gold nanoparticles with arm length 
of 130~nm and arm width of 50~nm, fabricated by electron-beam lithography on a glass substrate. The meta-atoms are distributed with a density of $6.5$ particles/$\mu \text{m}^2$ in a randomized spatial arrangement, which suppresses collective lattice resonances~\cite{BinAlam2021NatComm,Kolkowski2023APL}, while maintaining a strictly uniform global orientation $\theta$. The fabricated nanostructures were characterized by scanning electron microscopy (Fig.~\ref{fig:fab}a,b), revealing arm lengths and widths in good agreement with the target geometry, with slight deviations attributable to fabrication tolerances. The measured transmittance spectra (Fig.~\ref{fig:fab}c) confirm the absence of plasmonic resonances in the 900--970~nm pump wavelength range.

The three-bladed geometry was chosen for its $C_{3v}$ rotational symmetry ($m = 3$). From the selection rule of Eq.~(\ref{eq:selection}), the lowest allowed harmonic order is $n = jm - 1 = 2$ (with $j = 1$), confirming that SHG is the dominant nonlinear process. Furthermore, the $C_{3v}$ selection rules permit only the cross-polarized SHG channel, meaning the co-polarized channel is forbidden. Applying Eq.~(\ref{eq:PB_general}) with the $+$ sign (cross-polarized channel) and substituting $n = 2$, the nonlinear PB phase imparted to the SHG signal is:
\begin{equation}
\Phi_{\mathrm{PB}}= 3\sigma\theta,
\label{eq:PB_C3v}
\end{equation}
where $\sigma = \pm 1$ is the helicity of the incident pump. Importantly, the transmitted fundamental pump acquires no geometric phase upon rotation. This means that rotating the metasurface modifies only the SHG phase, leaving the pump unaffected. Furthermore, the threefold amplification in Eq.~(\ref{eq:PB_C3v}) means that a physical rotation of only $120^\circ$ spans the full $2\pi$ phase range.


\subsection{Experiments}   

\begin{figure} [t]
    \centering
    \includegraphics{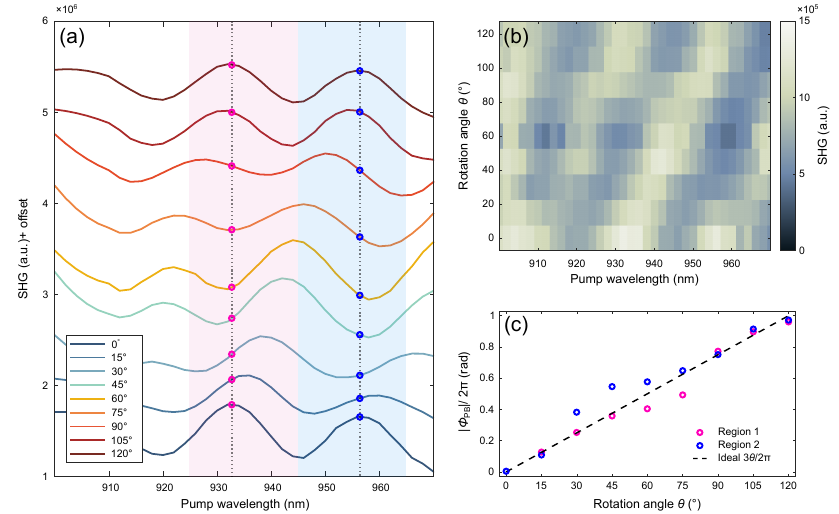}
    \caption{\textbf{SHG measurements and geometric phase extraction.} (a) SHG intensity spectra measured at different rotation angles $\theta$. As the rotation angle increases, the phase-matching peaks shift continuously across the spectrum. At $\theta=120^\circ$, each peak occupies the spectral position previously held by its neighbor, completing one full $2\pi$ geometric phase cycle. Spectra are vertically offset by $0.5 \times 10^6$~a.u.~for clarity. (b) Angle-resolved SHG map showing the SHG intensity (color scale) as a function of wavelength (x-axis) and rotation angle (y-axis). The diagonal shift of the intensity maxima confirms the continuous spectral tuning of the phase-matching peaks with rotation. (c) Geometric phase extracted from sinusoidal fits to the SHG intensity as a function of rotation angle, evaluated at the two representative wavelength regions highlighted in (a) (Peak 1, red circles; Peak 2, blue circles). The extracted phase increases linearly with rotation angle, following the ideal dependence (dashed black line), and spans a full $2\pi$ range upon a $120^{\circ}$ physical rotation, in good agreement with the theoretical prediction.}
    \label{fig:results}
\end{figure}
The experiment was performed using a femtosecond Titanium:Sapphire laser (Coherent Chameleon Vision II) as the fundamental pump, scanned across a spectral range of 900--970~nm with a pulse duration of around 200~fs and a repetition rate of 80~MHz. To control the SHG phase via the nonlinear PB phase, the pump was converted to a right-circularly polarized (RCP) state using a linear polarizer and a broadband zero-order quarter-wave plate. The average pump power incident on the first metasurface was maintained at 300~mW.

The MPC was constructed by placing two metasurface samples between two high-reflectivity plane mirrors (R~$>$~99\%), separated by a distance of $d$~=~3~cm. The mirrors were aligned to direct the beam through three passes per metasurface sample. This cascaded geometry establishes the interference-driven phase-matching condition responsible for the formation of discrete peaks in the SHG spectrum. A detailed description of the experimental setup is provided in the Supplementary Material. 

Both metasurfaces were initially co-aligned at a global orientation of $\theta = 0^\circ$. One metasurface remained fixed throughout, while the other was mounted on a motorized rotation stage and rotated in-plane from $0^\circ$ to $120^\circ$ in steps of $15^\circ$. Owing to the threefold symmetry of the three-bladed gold nanostructures, this $120^\circ$ physical rotation spans a complete $2\pi$ cycle of the nonlinear geometric phase. The transmitted SHG signal was collected and analyzed via a sensitive CMOS camera (ZWO-ASI1600MM).

As the geometric phase was tuned, the phase-matching peaks underwent a continuous spectral translation across the 900--970~nm pump range as shown in Fig.~\ref{fig:results}a. The angle-resolved SHG map confirms this behavior, showing a diagonal shift of the intensity maxima with increasing rotation angle (Fig.~\ref{fig:results}b). Upon reaching the full $120^\circ$ rotation, each peak had shifted to the spectral position previously occupied by its neighbor, completing one full $2\pi$ geometric phase cycle. To quantify this, we extracted the geometric phase from sinusoidal fits to the SHG intensity at two representative wavelength regions, finding linear growth with rotation angle in good agreement with the ideal $3\sigma\theta$ dependence (Fig.~\ref{fig:results}c). This deterministic and reversible spectral tuning demonstrates active control over the phase-matching condition 
of the SHG.

\section{Conclusions and outlook}
In summary, we have demonstrated active, mechanically reconfigurable control 
of SHG phase-matching conditions using $C_{3v}$-symmetric plasmonic metasurfaces integrated within an MPC. By introducing the nonlinear PB phase as a tunable bias in the phase-matching condition, we showed that rotating one metasurface relative to the other continuously shifts discrete phase-matching peaks across a 900--970~nm pump range. The threefold phase modification inherent to the $C_{3v}$ geometry enables a full $2\pi$ modulation with a $120^\circ$ physical rotation, and the extracted geometric phase shows good agreement with the ideal $3\sigma\theta$ dependence across the full tuning range. Crucially, this reconfigurability is achieved without modifying the nanostructure geometry, material composition, or pump conditions. 

The geometric phase control mechanism is decoupled from the SHG conversion efficiency itself, and could be combined with high-efficiency nonlinear sources in hybrid platforms. For instance, pairing a conventional SHG source---such as an ultra-thin slab of a nonlinear crystal---with a geometric-phase metasurface would provide additional dispersion control via resonance-based phase control~\cite{Almeida2016, Mekhael2026arXiv} or via the geometric phase as demonstrated here and in prior works~\cite{Ye2016NatCommun, Li2015NatMater, Liu2023ACSPhotonics, Tymchenko2015PRL, Liu2020AdvOptMater}. Such hybrid schemes could open new possibilities for nonlinear applications requiring broadband operation and higher-order dispersion management~\cite{Delong1994, Gallmann2000, Konishi2020APL}, combining ultrathin nonlinear materials with the continuous, broadband, and post-fabrication reconfigurability of the geometric phase.

A practical route toward increased efficiency would be to replace one or both mirrors of the MPC with a reflective metasurface---for instance, a resonant waveguide grating designed to simultaneously act as a high-reflectivity mirror and a nonlinear geometric-phase element~\cite{Saari2010, Siltanen2007, Fehrembach2021, Quaranta2018}---thereby achieving a more compact geometry with reduced propagation losses. More broadly, these results establish engineered rotational symmetry in plasmonic meta-atoms as a practical design principle for reconfigurable nonlinear flat optics, opening new possibilities for tunable frequency conversion and nonlinear wavefront shaping in compact, ultrathin optical platforms.

\section*{Acknowledgments}
The authors acknowledge fruitful discussions with Thomas Zentgraf. This work was supported by Research Council of Finland Flagship Programme, Photonics Research and Innovation PREIN 346518 and Research Council of Finland project number 359450.

\bibliographystyle{IEEEtran}
\bibliography{references}
\end{document}


\maketitle
\section*{Nonlinear Optical Measurements}
\begin{figure}
    \centering
    \includegraphics{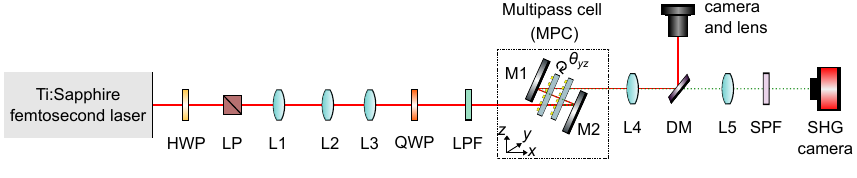}
    \caption{\textbf{Experimental setup.} Schematic of the setup used for SHG measurements from a dual-metasurface multipass cell at different rotation angles of one metasurface.}
    \label{fig:setup}
\end{figure}

The experimental setup illustrated in Fig.~\ref{fig:setup} was used to characterize the second-harmonic generation (SHG) from the metasurfaces. A tunable titanium:sapphire femtosecond laser (Coherent Chameleon Vision II), operating at a repetition rate of 80~MHz with a spectral range of 680--1080~nm and a pulse duration of around 200~fs, served as the pump source. The pump wavelength was scanned across the 900--970~nm range in steps of 2~nm at each rotation angle. A half-wave plate (HWP) combined with a linear polarizer (LP) was used to control the pump power incident on the sample. Three lenses L1 ($f_1 = 75$~mm), L2 ($f_2 = 25.4$~mm), and L3 ($f_3 = 250$~mm) were used to adjust the beam diameter and set the focusing conditions. A quarter-wave plate (QWP) placed after L3 converted the linearly polarized beam to right-circular polarization (RCP), and a long-pass filter (LPF) was used to suppress any residual wavelengths below the pump range before the beam entered the multipass cell.

The multipass cell (MPC) consisted of two high-reflectivity mirrors M1 and M2 (R~$>$~99\%) enclosing two metasurface samples. The mirrors were aligned such that the pump beam interacted three times with each metasurface sample, establishing the phase-matching condition responsible for the peaks in the SHG spectrum. The first metasurface was mounted on a motorized rotation stage, allowing in-plane rotation by an angle $\theta_{yz}$ about the beam propagation axis ($x$). The beam waist at this metasurface was approximately 340~$\mu$m, corresponding to a Rayleigh range of 9.75~cm, ensuring uniform interaction with both metasurfaces throughout the multipass cell.

After the MPC, lens L4 ($f_4 = 75$~mm) collected and collimated the output beam. A dichroic mirror (DM) separated the SHG signal from the residual pump, directing the reflected pump beam to a camera for sample imaging and beam position control. The transmitted SHG signal was then focused by lens L5 ($f_5 = 75$~mm) and further filtered by a short-pass filter (SPF) to suppress any remaining pump light, before being imaged onto a sensitive CMOS camera (ZWO-ASI1600MM).
